# Extension of the Planar Noh Problem to Aluminum, Iron, Copper, and Tungsten

Chloe' E. Yorke, April D. Howard, Sarah C. Burnett, Kevin G. Honnell[a], Scott D. Ramsey, and Robert L. Singleton, Jr.

*Computational Physics & Theoretical Design Divisions*
*Los Alamos National Laboratory*
*Los Alamos, NM  USA*

[a] Corresponding author: kgh@lanl.gov

**Abstract.** The classic Noh verification test problem is extended beyond the traditional ideal gas and applied to shock compression of condensed matter. Using the stiff-gas equation of state (EOS), which admits an exact analytical solution for the planar Noh problem, we examine the shock compression of Al, Fe, Cu, and W. Analytical EOS predictions for the jump in density and the location of the shock are compared to numerical results obtained using the same EOS within Los Alamos compressible-flow codes Flag and xRage. Excellent agreement between the numerical and exact results is observed. Both codes exhibit first-order spatial convergence with increasing mesh resolution.

## I. INTRODUCTION

The "Noh problem" is classic verification problem in the field of computational hydrodynamics of compressible flows.[1-7] A strong outward facing shockwave is created by impinging a uniformly flowing ideal gas on a hard wall (or imploding it onto the axis of a cylinder or the center of a sphere, depending on the geometry of interest); see Fig. 1. A simple problem to conceptualize and one that admits an exact analytical solution,[7] it is nonetheless difficult for numerical codes to predict correctly, making it an ideal code-verification test bed. In its original incarnation,[1] and in nearly all applications since, the fluid is a simple ideal gas initialized at the rather unrealistic conditions of zero temperature, energy, and pressure;[2-4] once verified, however, these codes are often used to study highly non-ideal fluids and solids under conditions far removed from zero temperature and pressure. It would seem advantageous, then, to develop complementary code verification test problems that probe code behavior under more realistic conditions.

In this work we describe how the canonical planar Noh problem may be extended beyond the commonly studied polytropic ideal gas to more realistic conditions and materials by using the stiff-gas equation of state (EOS).[5-12] The stiff-gas EOS retains much of the simplicity of the ideal gas while providing a qualitatively accurate representation of the shock compression of condensed matter. Moreover, it admits an analytical solution to the planar Noh problem, creating a means of performing code verification, validation (V&V), and uncertainty quantification (UQ) under physically realistic conditions.

## II. THEORY

### A. Stiff-Gas Equation of State

The stiff-gas EOS[5-12] relates the pressure, $P$, to the density, $\rho$, and specific internal energy, $E$:

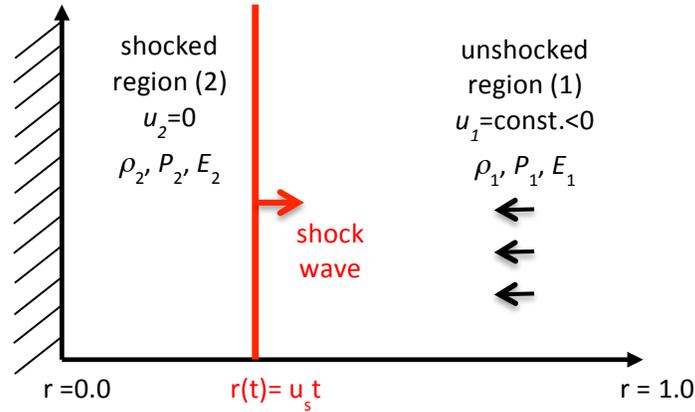

**FIGURE 1.** The planar Noh problem: A uniform, inwardly flowing fluid impinges on a hard wall at $t=0$, setting up an outwardly traveling shock wave moving at a constant velocity $u_s$ with a stagnant fluid behind the shock front.

$$P(\rho,E) = (\rho - \rho_o)c_s^2 + \rho(\gamma - 1)E \quad (1)$$

where $\rho_o$ and $c_s$ denote the density and sound speed at ambient temperature and pressure, and $\gamma$ is an adjustable parameter akin the to adiabatic index of an ideal gas. Eq. (1) can be considered as a simple modification of the ideal gas in which the pressure and energy have been shifted upward by $\rho_o c_s^2$ and $c_s^2/(\gamma -1)$ respectively; alternatively, Eq. 1 may be viewed as simplified Grüneisen EOS[13] based off of a Taylor expansion of the principal isentrope, with $\gamma-1$ functioning as the Grüneisen parameter. Eq. (1) retains much of the mathematical simplicity of the ideal gas while providing a qualitatively accurate description of real condensed matter, including a realistic bulk modulus (via $c_s$) and the ability to go into tension when $\rho < \rho_o$.

## B. Analytical Solution for the Planar Noh Problem

Using Lie group methods, Axford[5] and more recently, Ramsey, Burnett, et al.[6,7] constructed analytical solutions to the Noh problem for several non-ideal EOSs, including the stiff gas. A similarity transform reduces the Euler equations to three coupled ordinary differential equations for mass, momentum, and energy conservation, which in combination with the Noh boundary and initial conditions and the Rankine-Hugoniot[13] jump conditions yield:[7]

$$u_s = a/(4\sqrt{\rho_1}) \quad (2)$$

$$\rho_2 = \rho_1(1 + 4\sqrt{\rho_1}|u_1|/a) \quad (3)$$

$$P_2 = P_1 + \rho_2 |u_1| u_s \quad (4)$$

where $u_s$ denotes the shock speed, $u_1$ the initial velocity, subscripts 1 and 2 refer to unshocked and shocked fluid regions, and

$$a = \sqrt{16(\rho_o c_s^2 + \gamma P_1) + (\gamma+1)^2 \rho_1 u_1^2} - (3-\gamma)\sqrt{\rho_1}|u_1| \quad (5)$$

## III. VERIFICATION STUDIES

We extend the Noh test problem to Al, Fe, Cu, and W by employing Eq. (1) in hydrocode simulations in place of the ideal gas EOS, using experimentally-determined values for $\rho_o$ and $c_s$ (Table 1).[12,14] Effective $\gamma$ parameters were empirically obtained by combining Eq. (1) with the Rankine-Hugoniot relation[13] and fitting it to the experimental principal Hugoniots of each material.[15] Reasonably good fits were obtained in all four cases, as shown in Fig. 2 for

W. Initial velocities, $u_1$, were selected by choosing a range of desired shock pressures, $P_2$, that spanned the experimental data and then back-calculating the required $u_1$ from Eq. (4) in combination with (2), (3), and (5).

**TABLE 1.** Parameters used in the stiff-gas EOS (Eq. 1) to model shock compression of Al, Fe, Cu, and W.[12,14]

| Metal | $\rho_\circ$ (g/cm³) | $c_s$ (km/s) | $\gamma$ | $|u_1|$ (km/s) |
|---|---|---|---|---|
| Al | 2.784 | 5.328 | 3.05 | 0.41 – 2.6 |
| Cu | 8.924 | 3.94 | 3.53 | 0.03 – 2.8 |
| Fe | 7.856 | 3.80 | 3.83 | 0.07 – 3.2 |
| W | 19.235 | 4.029 | 2.99 | 0.16 – 2.1 |

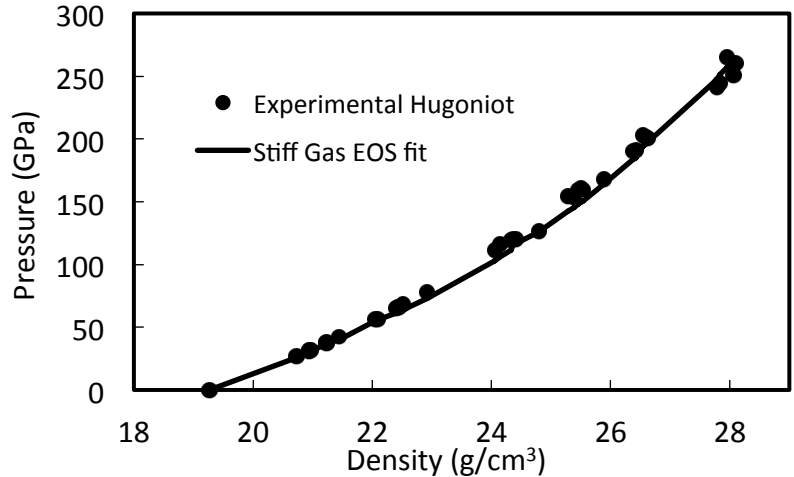

**FIGURE 2.** Comparison of the stiff-gas EOS prediction for the principal Hugoniot of W, using $\gamma = 2.99$, to experimental shock data. Similarly good fits were obtained for Al, Fe, and Cu using parameters reported in Table 1.

Simulations were conducted using the Los Alamos Lagrangian code Flag[16] and the Eulearian code xRage,[17] initialized at ambient pressure and density (i.e., $P_1 = 1$ atm., $\rho_1 = \rho_o$). Representative results for the density as a function of distance from the wall are shown for W in Fig. 3, using an initial velocity of 2.1 km/s and four mesh sizes. Code predictions for the pressure are compared with exact result in Fig. 4. Simulations shown in Fig. 3 correspond to the highest pressure point in Fig. 4 (261 GPa). In the sixteen test cases studied (four materials at four impact velocities), both codes accurately predicted the location and magnitude of the pressure, energy, and density jumps across the shock; however the usual errors[1-4] associated with the use of artificial viscosity immediately adjacent to the wall (the so-called "wall heating" effect) and at the shock front were observed, though their magnitude as measured by the L1 norm was considerably smaller than those seen in canonical ideal-gas case.[7] Mesh convergence studies, using the V&V tool ExactPack,[18] indicated first-order convergence of these errors with increasing mesh resolution (Fig. 5).

## IV. CONCLUSION

By employing mathematically simple but qualitatively realistic EOSs, the domain of code verification test problems can be extended from ideal gases to situations more closely resembling actual applications. In this study we applied the stiff-gas EOS to the shock compression of Al, Fe, Cu, and W. Comparison of exact predictions for the planar Noh problem to numerical results for two compressible-flow hydrocodes, Flag and xRage, showed excellent agreement. While errors due to wall heating were observed, they were considerably smaller than those seen in the canonical ideal-gas case.[7] It would be of interest to explore the extension of this work to other classic verification problems, such as the Sod[3,18,19] and Sedov[18,20] problems.

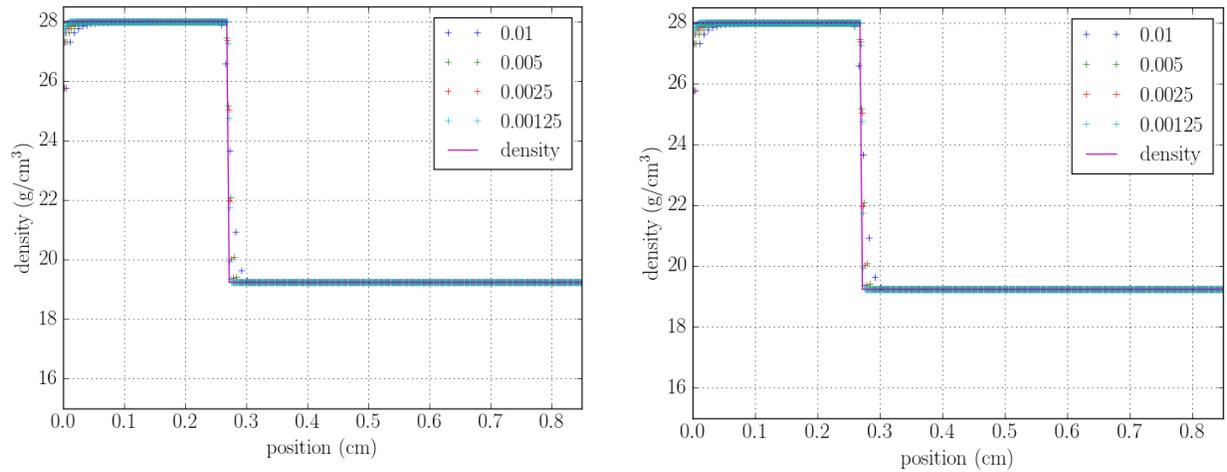

**FIGURE 3.** Density as a function of position for W at 28 GPa. Points represent simulation results from the Lagrangian code Flag (left) and the Eulearian code xRage (right) as a function of zone sizes (cm); solid lines are the exact stiff –gas response.

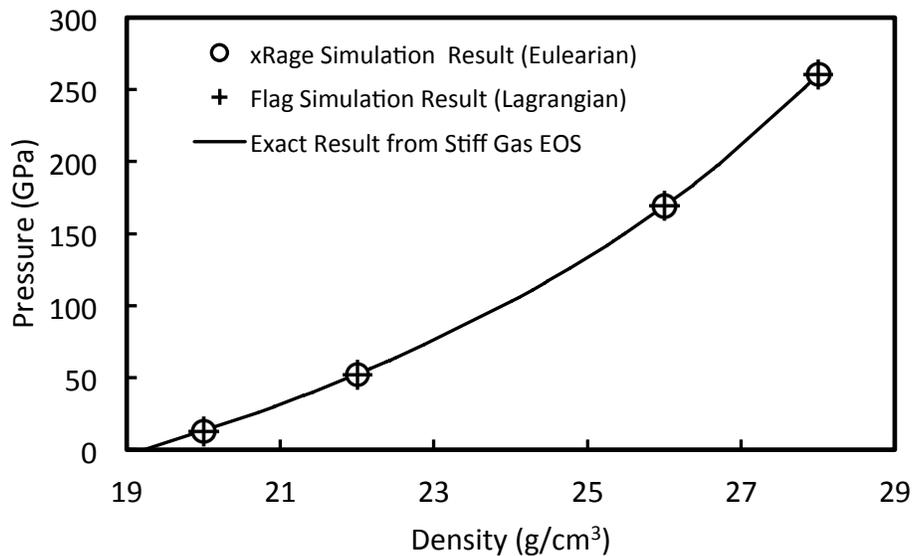

**FIGURE 4.** Principal Hugoniot of W. Points denote the simulations predictions at four pressures using Flag and xRage in combination with the stiff-gas EOS; solid line is the exact stiff-gas result.

## ACKNOWLEDGEMENT

The authors gratefully acknowledge the support of the U.S. DOE NNSA Minority Serving Institution Program and the Advanced Strategic Computing Program, in particular the V&V and Physics and Engineering Models projects. Los Alamos National Laboratory is operated on behalf of the NNSA by Los Alamos National Security, LLC, under Contract No. DE-AC52-06NA25396.

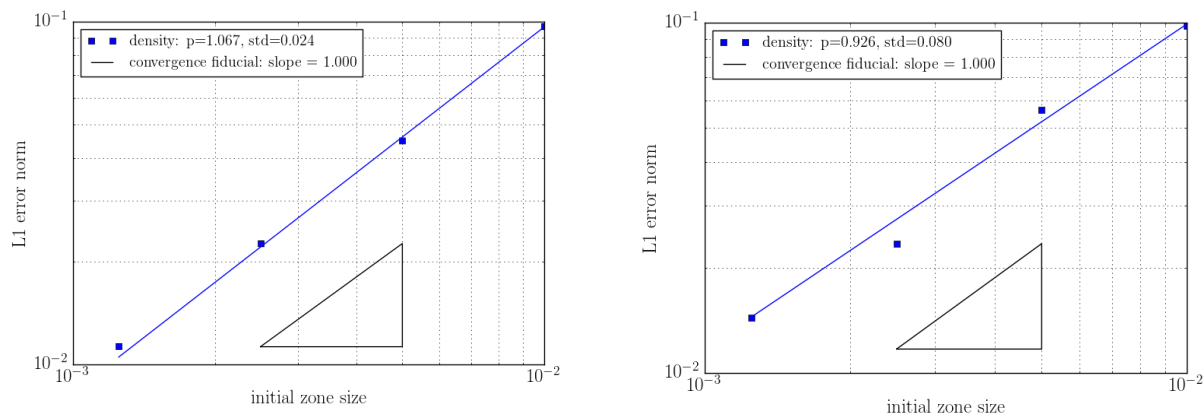

**FIGURE 5.** Spatial convergence studies in Flag (left) and xRage (right) for the simulations shown in Fig. 3. Points represent the numerical errors associated with a particular zone size (cm), as reflected by the L1 norm. The triangles are fiducial guides of unit slope. Both codes exhibit approximately first-order convergence.

# REFERENCES

1. W. F. Noh, J. Comput. Phys. **72,** 78–120 (1987).
2. M. Gehmeyr, B. Cheng, and D. Mihalas, Shock Waves **7**, 255–274 (1997).
3. J. Campbell and R. Vignjevic, "Artificial Viscosity Methods for Modeling Shock Wave Propagation," in *Predictive Modeling of Dynamic Processes*, edited by K. Thoma (Springer, Dordrecht, 2009), pp. 349-366.
4. W. J. Rider, J. Comput. Phys. **162** 395–410 (2000).
5. R. A. Axford, Laser Part. Beams **18,** 93–100 (2000).
6. S. D. Ramsey, Z. M. Boyd, and S. C. Burnett, Shock Waves **27**, 477-485 (2017).
7. S. C. Burnett, K. G. Honnell, S. C. Ramsey, S. D. Ramsey, and R. L. Singleton, Jr., J. Verification, Validation, and Uncertainty Quantification (submitted).
8. F. H. Harlow and W. E. Pracht, Phys. Fluids **9**, 1951-1959 (1966).
9. Y. S. Wei and R. J. Sadus, AIChE J. **46**, 169-196 (2000).
10. R. H. Cole, *Underwater Explosions* (Princeton U. Press, Princeton, 1948).
11. R. Courant, *Supersonic Flow and Shock Waves: A Manual on the Mathematical Theory of Non-linear Wave Motion* (Courant Institute, NYU, New York, 1944).
12. L. Weixin, "Simplified Equation of State P = P($\rho$,E) and P=P($\rho$,T) for Condensed Matter," in *Shock Waves in Condensed Matter*, edited by Y. M. Gupta (Plenum, NY, 1986), pp. 167-173.
13. S. Eliezer, A. Ghatak, and H. Hora, *Fundamentals of Equations of* State (World Scientific, Singapore, 2002), pp. 153-164, 176-183, 200-204.
14. R. G. McQueen, S. P. Marsh, J. W. Taylor, NJ. N. Fritz, and W. J. Carter, "The Equation of State of Solids from Shock Wave Studies," in *High Velocity Impact Phenomena*, edited by R. Kinslow (Academic, NY, 1970), pp. 294-419.
15. S.P. Marsh, editor, *LASL Shock Hugoniot Data* (Univ. of Calif. Press, Berkeley,1980).
16. D. E. Burton, "Flag, Multi-Dimensional Hydrodynamics using Polyhedral Grids," LANL report no. LA-UR-97-1109 (Los Alamos National Laboratory, Los Alamos, NM, 1997).
17. M. Gittings, R. Weaver, M. Clover, T. Betlach, N. Byrne, R. Coker, E. Dendy, R. Hueckstaedt, K. New, W.R. Oakes, D. Ranta, and R. Stefan, Comp. Sci. Disc. **1**, 015005 (2008).
18. R. L. Singleton, Jr., D. M. Israel, S. W. Doebling, C. N. Woods, A. Kaul, J. W. Walter, Jr., M. L. Rogers, "ExactPack Documentation", LANL report no. LA-UR-16-23260 (Los Alamos National Laboratory, Los Alamos, NM, 2016).
19. L. I. Sedov, *Similarity and Dimensional Methods in Mechanics* (Academic, New York, 1959), pp. 146-304.
20. G. A. Sod, J. Comput. Phys. **27**, 1-31 (1978).